\documentclass[%
 reprint,showkey,superscriptaddress,
 amsmath,amssymb,
 aps,
]{revtex4-2}

\usepackage{graphicx}
\usepackage{dcolumn}
\usepackage{bm}

\begin{document}


\title{Spatial structure and information transfer in visual networks}

\author{Winnie Poel}\email{winnie.poel@gmail.com}
\affiliation{Institute for Theoretical Biology, Department of Biology, Humboldt Universit{\"a}t zu Berlin, D-10099 Berlin, Germany }%
 \affiliation{Bernstein Center for Computational Neuroscience Berlin, D-10115 Berlin, Germany }
 
\author{Claudia Winklmayr}
 \affiliation{Bernstein Center for Computational Neuroscience Berlin, D-10115 Berlin, Germany }

\author{Pawel Romanczuk}
\email{pawel.romanczuk@hu-berlin.de}
\affiliation{Institute for Theoretical Biology, Department of Biology, Humboldt Universit{\"a}t zu Berlin, D-10099 Berlin, Germany }%
 \affiliation{Bernstein Center for Computational Neuroscience Berlin, D-10115 Berlin, Germany }

\date{\today}

\begin{abstract}
In human and animal groups, social interactions often rely on the transmission of information via visual observation of the behavior of others.  These visual interactions are governed by the laws of physics and sensory limits. Individuals appear smaller when far away and thus become harder to detect visually, while close by neighbors tend to occlude large areas of the visual field and block out interactions with individuals behind them. Here, we systematically study the effect of a group's spatial structure, its density as well as polarization  and aspect ratio of the physical bodies, on the properties of the visual interaction network.
In such a network individuals are connected if they can see each other as opposed to other interaction models such as metric or topological networks that omit these limitations due to the individual's physical bodies. We study the effect that spatial configuration has on the static properties of these networks as well as its influence on the transmission of information or behaviors which we investigate via two generic models of social contagion.
We expect our work to have implications for the study of animal groups, where it could inform the study of functional benefits of different macroscopic states. It may also be applicable to the construction of robotic swarms communicating via vision or for understanding the spread of panics in human crowds.
\end{abstract}

\maketitle

\section{Introduction}
The emergent collective behavior of animal groups, or more generally multi-agent systems, is decisively shaped by the structural properties of the underlying networks of social interactions \cite{ballerini2008interaction,strandburg2013visual,rahmani2020flocking}. These networks may strongly differ in in their spatio-temporal embedding depending on the type of interaction or the behavior of interest. For example in humans, online social networks have no, or only very weak, relation to physical space and interactions typically do not depend on instantaneous communication \cite{kumar2010structure,mislove2007measurement}. On the other hand, contact networks governing the (direct) spread of pathogens between individuals  \cite{balcan2009multiscale,isella2011s} or interaction networks governing the collective movement of human crowds  \cite{Moussaid2011Pedestrian,moussaid2016crowd,wirth2016visual}, represent examples of spatial networks with a tight correlation between physical distance of individuals and the probability (or strength) of corresponding interaction links. In non-human animals a similar variety of networks can be observed ranging from mating and hierarchy networks \cite{krause2015animal} to strongly spatio-temporally constrained interaction networks underlying the collective movement of fish schools \cite{strandburg2013visual,Rosenthal2015}, bird flocks \cite{ballerini2008interaction,ling2019behavioural} or insect swarms and colonies \cite{sarfati2020spatio,wild2021social}. Especially in large animal collectives many inter-individual interactions forming the basis for coordinated collective movements \cite{vicsek2012collective}, collective decision making \cite{couzin2005effective}, or spread of information \cite{Rosenthal2015} are directly governed by spatial proximity.

Typically, spatially embedded interaction networks between biological agents are modelled either via {\em metric} network models {\cite{couzin2005effective}}, where the probability of a link (or its strength) depends only on the inter-individual distance, or by {\em topological} models \cite{rahmani2020flocking} where a focal agent is connected to a set of spatial neighbors with a high closeness rank but where the actual link probability (or strength) does not dependent on the actual physical distance as e.g. in the $k$-nearest neighbor network model \cite{ballerini2008interaction,rahmani2020flocking}. In the past, most agent-based models assumed metric interaction networks, but after evidence for topological interaction in starling flocks has been presented by \cite{ballerini2008interaction}, corresponding topological interaction networks have received increased attention in the context of collective animal behavior.

However, the discussion of these two idealized models of interaction networks largely ignores the constraints set by different sensory and cognitive mechanisms underlying social interactions \cite[see e.g.][]{lemasson2009collective}. Vision mediated interactions play an important role for a wide range of social phenomena \cite{Moussaid2011Pedestrian,strandburg2013visual,SosTwoBak19, Bastien2020VisualModel}. In particular, visual interaction networks accounting for visual occlusions have been shown to outperform both metric and topological interaction networks in describing collective behavior of fish \cite{strandburg2013visual}. Acoustic communication, on the other hand,  which shapes social behavior of many animals \cite{fichtel2010vocal,demartsev2018vocal}, is not affected by the same constraints as vision (e.g occlusion at high densities) but depends mainly on sensory limits and properties of the medium.
Here, metric interaction networks may provide a simple model for acoustic social interactions. Finally, topological interaction networks with a limited number of nearest neighbors have been recently  discussed in the context of cognitive constraints regarding the number of neighbors (or objects) a focal individual can pay attention to \cite{rahmani2020flocking}. 

Although the importance of visual interactions has been highlighted in recent research  \cite{strandburg2013visual,Rosenthal2015}, there is a lack of systematic investigations of the structure of visual networks, in particular with respect to their ability to transmit information and behavior.
Here, we address this gap by comparing visual networks with the established metric and topological models which, as discussed above, may represent different sensory and/or cognitive constraints.

Comparing different types of networks quantitatively is challenging and even more so when networks represent social interactions based on different sensory limits which will be unique for a given biological agent and environment and can not generally be related to each other (i.e. some species may be able to hear further than they can see while for others the opposite may be true). While a common approach is to quantitatively compare networks of similar average degree (i.e. average number of interaction partners per individual) this may not yield the most relevant insights into biological systems where sensory limits may be fixed and tuning them to achieve a certain number of interactions may not be possible. It is known though that animal groups can quickly modify their spatial density of individuals in response to changes in the environment, e.g. related to predation risk \cite{SosTwoBak19,romenskyy2020quantifying} which in turn influences network structure. Thus besides the common approach of quantitatively comparing networks of similar degree, we also especially focus on a qualitative comparison of the networks' dependence on the spatial density of individuals.

First, we study structural differences between the three network types using static network measures that have been used to classify and compare different types of networks \cite{newman2003structure,Barthelemy2018transitions}. We then move beyond pure structural network analysis and compare the dynamics of a simple and a complex contagion process on visual and metric networks. While simple contagion may be viewed as a minimal model for information diffusion in animal interaction networks, complex contagion describes the spread of behaviors or emotions where simultaneous, non-linear reinforcement by multiple neighbors is at play.

Our work demonstrate the fundamental difference of visual networks in terms of structural parameters in comparison to metric and topological interaction networks. In particular, visual networks exhibit qualitatively different behavior in response to density modulation. Second, we demonstrate that these structural differences result in characteristic deviations in the dynamics of the two contagion processes on visual and metric networks.

\section{Methods}
In order to investigate and compare the influence of spatial structure on properties of and dynamics on visual networks we will construct  visual, metric and topological networks in two dimensional space. Here, we first discuss how we generate and characterize the spatial distribution of individuals, i.e. the network nodes. We then move on to explain the construction of the different spatially-embedded networks and the network measures we will use to characterize them, before introducing the two contagion models that we use to investigate the transfer of information or behaviors on these networks.

\subsection{Network construction: Spatial distribution and shape of nodes}
We initially place $N=n^2$ nodes on an $n$ by $n$ two dimensional square grid with distance $g$ between nearest neighbors. This setup creates a homogeneous density controlled via the parameter $g$. To yield more realistic distributions we add positional noise, $\eta_x,\eta_y \sim\mathcal{U}(-\eta_{pos},\eta_{pos})$, drawn from the uniform random distribution between $\pm \eta_{pos}$ which we also scale by a factor $g$. Thus, node positions are given by
\begin{equation}
    \vec{r}_{i}=\left(\begin{array}{c} x_{i}\\y_{i} \end{array}\right) =g\left(\begin{array}{c} n_x+\eta_x\\n_y+\eta_y\end{array}\right)
\end{equation}
where $n_x, n_y \in (0,1,\dots,n-1)$ with $i=n n_x+n_y$. Here, we only consider values of $\eta_{pos}$ up to 0.5 to ensure that the density stays relatively homogeneous which allows us to systematically study its effect on the networks by varying $g$. Note that in the limit of large noise $\eta_{pos}$, the random placement of agents will correspond to the simple two-dimensional spatial Poisson process.

Throughout this work we mainly characterize spatial distribution of nodes by density and polarization. We estimate spatial density, $\rho$, via the average third nearest neighbor distance, $\bar{r}_{3nnd}$, a measure which describes the average radius of a disk containing four individuals (the focal individual and its three nearest neighbors, for a sketch see appendix). An estimate of the local density which is relatively robust with respect to positional noise is thus given by
\begin{equation}
    \rho=\frac{4}{\pi \bar{r}_{3nnd}^2}~.
\end{equation}

Inspired by the elongated body shape of fish, single agents are represented by identical ellipses of length 1 and width $w$ with orientations $\phi_i$ that are drawn from a von Mises distribution
\begin{equation}
    f(\phi|\mu,\kappa)=\frac{e^{\kappa\cos(\phi-\mu)}}{2\pi I_0(\kappa)}
\end{equation}
where $\mu=0$ is the average, $I_0(\kappa)$ is the modified Bessel function of order 0 and $\kappa$ a parameter that defines the width of the distribution (here we use $\kappa=0.1, 1.7, 31.6$). Note that because ellipses are of length one, all units can be understood as measured in body length (BL).
 We characterize the group's \textit{polarization} (degree of orientational order) by the absolute value of the normalized sum of all orientation unit vectors $\vec{\phi_i}=\left(\cos(\phi_i), \sin(\phi_i)\right)^T$
\begin{equation}
    \Phi=\frac{1}{N}\left|\sum_{i=1}^N\vec{\phi_i}\right|~.
\end{equation}
To eliminate overlap of the ellipses (to keep the group strictly two dimensional) we use simulations of ellipse shaped particles based on the code provided by \cite{Palachanis2015}. These particles repel each other with a force proportional to their overlap area and we let them settle into a non-overlapping configuration (see appendix). Thus there is a upper limit to density given by the physical bodies of the nodes. An example of a spatial configuration generated using $g=1.4, \eta_{pos}=0.5, N=36, \kappa=0.9, w=0.3$ is shown in Figure \ref{fig:networks}B.

\subsection{Network construction: Edges}
The three network types considered in this work are distinguished exclusively via the rules for the construction of links based on spatial positions of the ellipse shaped bodies, i.e. the network nodes. We limit ourselves to binary networks with the adjacency matrix given by
\begin{equation}
    A_{ij}=\left\{\begin{array}{ll}1& \text{if there is a link from $i$ to $j$}\\0& \text{otherwise}\end{array}\right.~.
\end{equation}
The decision rule determining if a link from $i$ to $j$ exists depends on the type of network and is explained in the following paragraphs. An illustration of the connection rules can be seen in Figure \ref{fig:networks}A. Figure \ref{fig:networks} provides an example of each network type for $N=36$, where the same positions and orientations of ellipses are underlying each network and respective thresholds are chosen such that the total number of links is identical between networks.

\subsubsection{Visual networks}
In a visual network a link from node $i$ to node $j$ exists if $i$ is visible to $j$. For simplicity, we assume each ellipse has $360^\circ$ vision from a single eye located at the center of the ellipse. We determine the angle, $\alpha_{ij}$ of an ellipse $i$ in the visual field of $j$ through a combination of analytical calculation of the (unobstructed) visual angle, $\alpha^\text{max}_{ij}\geq\alpha_{ij}$, with a numerical casting of rays at angles specified by the analytics to determine occlusions (for details refer to the appendix and \cite{LeBlanc2018Thesis}). We will further refer to this visual angle as the \textit{angular area} of ellipse $i$ for the focal individual $j$. Figure \ref{fig:networks}A illustrates the visual field of a central ellipse as shaded areas. To account for sensory and/or cognitive limitations an individual $i$ must haven an angular area above a certain threshold, $\theta_\text{visual}$, in order to be visible to $j$. This threshold parameterizes our visual network model
\begin{equation}
    A^\text{visual}_{ij}=\left\{\begin{array}{ll}1& \text{if }\alpha_{ij}\geq\theta_\text{visual}~\\
    0 & \text{otherwise}\end{array}\right.~.
\end{equation}
In Figure \ref{fig:networks}A angular areas larger than the visual threshold, $\theta_\text{visual}=0.29$, are indicated by dark grey areas while those smaller than the visual threshold are shaded light grey. Correspondingly, it can be seen that the central individual only has incoming links from those individuals with $\alpha_{ij}\geq\theta_\text{visual}$. Because visibility is not necessarily reciprocal, visual networks are generally directed.

\subsubsection{Metric networks}
In a metric network two nodes are connected if their Euclidian distance, $r_{ij}=|\vec{r_i}-\vec{r_j}|$, is smaller than the metric threshold, $\theta_\text{metric}$ \cite{Barthelemy2011spatial}
\begin{equation}
    A^\text{metric}_{ij}=\left\{\begin{array}{ll}1& \text{if }r_{ij}\leq\theta_\text{metric}~\\
    0 & \text{otherwise}\end{array}\right.~.
\end{equation}
This rule is illustrated in Figure \ref{fig:networks}A, where a focal individual (white) has incoming links from all individuals within a radius of $\theta_\text{metric}$ (marked by a shaded circle). Because $r_{ij}=r_{ji}$ metric networks are undirected. 
Note that, when not considering occlusion, visual networks can also be understood as having a metric interaction range because the visual threshold sets an upper limit for the interaction distance. The exact distance at which the angular area of an ellipse drops below the visual threshold, will of course depend on its orientation and width (see appendix). When needing to construct the visual threshold resulting in an equivalent effective cutoff distance for visual networks as in a certain metric network, we therefore averaged over all possible relative orientations for a specific value of $w$ (see appendix).

\subsubsection{Topological networks}
In a topological network node $i$ has incoming links from its $\theta_\text{topo}$ nearest neighbors, chosen successively by increasing Euclidian distance. If we assign each individual $j$ a closeness rank, $k_{ij}$, with respect to individual $i$ as $k_{ij}=\big|\{m | r_{im}<r_{ij}\}\big|+1$ where $||$ denotes the set's cardinality (number of elements in the set) we can write the construction rule as
\begin{equation}
    A^\text{topo}_{ij}=\left\{\begin{array}{ll}1& \text{if }k_{ij}\leq\theta_\text{topo}~\\
    0 & \text{otherwise}\end{array}\right.~.
\end{equation}

The connection rule is illustrated in Figure \ref{fig:networks}A, where individuals close to a focal individual (white) are labeled according to their closeness rank and the focal individual has incoming links from those with a closeness-rank up to $\theta_\text{topo}=2$. Because closeness rank is not necessarily reciprocal, topological networks are generally directed as can been seen in the example in Figure \ref{fig:networks}B.
\subsection{Network measures}
To assess and compare the structural properties of the different networks we use three well-established measures, the average in-degree, the average clustering coefficient and average shortest path length. These measures have been used widely to classify and compare different types of networks (see e.g \cite{newman2003structure,Barthelemy2011spatial}). We calculate the latter two quantities using the methods implemented in Python's \texttt{networkx} library \cite{SciPyProceedings_11}. 

\subsubsection{In-degree and Out-degree}
The in-degree $d_i^{in}$ of a node $i$ is defined as the number of its incoming links, \begin{equation}
d^{in}_i=\sum_{j\neq i}A_{ji}~.
\end{equation} 
The average in-degree is given by $\bar{d}^{in}=\frac{1}{N}\sum_{i=1}^Nd^{in}_i$. The out-degree is analogously defined as
\begin{equation}d_i^{out}=\sum_{j\neq i}A_{ij}\end{equation}. 

\subsubsection{Clustering coefficient}
In an undirected, unweighted network the clustering coefficient of node $i$ describes the probability that two neighbors, $j$ and $k$ of node $i$ are also linked among each other. This is calculated by dividing the number $t_i$ of all triangles formed by $i$ by the number $T_i$ of all possible triangles that could be formed by $i$. Here, we use the following simple extension of this measure to directed graphs from \cite{Fagiolo2007Clustering}
\begin{equation}
\begin{aligned}
    c_i&=\frac{t_i}{T_i}\\
    &=\frac{\frac{1}{2}\sum_{j\neq i}\sum_{k\neq(i,j)}(A_{ij}+A_{ji})(A_{ik}+A_{ki})(A_{jk}+A_{kj})}{d_i^{tot}(d^{tot}_i-1)-2d^\leftrightarrow_i}\\
    d^{tot}_i&=d_i^{in}+d_i^{out},\quad
    d_i^\leftrightarrow=\sum_{i\neq j}A_{ij}A_{ji}~.
\end{aligned}
\end{equation}
The average clustering coefficient of a network is then given by $\frac{1}{N}\sum_{i=1}^Nc_i$. 

\subsubsection{Average shortest path length}
The average shortest path length describes the average minimum number of steps on the network needed to get from a node $i$ to a node $j$. It is defined as
\begin{equation}
    a=\frac{1}{N(N-1)}\sum_{i, j} d(i,j)
\end{equation}
where $d(i,j)$ is the length of the shortest path between nodes $i$ and $j$. We use \texttt{networkx}'s implemented algorithm \texttt{average\_shortest\_path\_length} to determine the value for $a$ where possible (networks need to be weakly connected).
\subsubsection{Relative link length}
While the shortest path measures the topological distance between nodes in spatial networks the link length, ${l}_{ij}=r_{ij} \ \text{if}\  A_{ij}=1$, measures the Euclidean distance between two connected agents. In order to make link length comparable across densities, we measure link length in units of the longest possible link, $l^\text{max}$, in the group. For the grid configuration used in this paper this \textit{relative link length} is given by 
\begin{equation}
    l_{ij}^\text{rel}=\frac{l_{ij}}{l^\text{max}}=\frac{r_{ij}}{g\sqrt{2}(\sqrt{N}-1)}~.
\end{equation}

\subsection{Contagion models}
We investigate two models, one of simple and one of complex contagion, to demonstrate the differential impact of the network topology on these processes. In simple contagion, the probability of an infection in a time interval $\Delta t$ can be decomposed into the superposition of independent pair-wise interactions between a non-infected (susceptible) individual and its infected (network) neighbors. In complex contagion such a decomposition is not possible as the infection probability is a non-linear function of the number or fraction of infected neighbors. 
We emphasize that 'infection' does not refer here to disease spread, but to spreading of information or behavior. Thus, throughout this work, becoming infected refers to an individual becoming informed or activated \cite[see e.g.][]{SosTwoBak19}.
Both types of contagion models are studied via discrete time approximation of the continuous time stochastic infection and recovery processes using a (small) numerical time step $\Delta t=0.05$. In what follows we describe the respective processes.

\subsubsection{Simple contagion model} 
Each agent within the network can be in one of three states: susceptible S, infected I or recovered R. A susceptible agent in contact with a single infected neighbor can become infected with a constant probability rate $\beta$. Thus, the infection probability for such a pair-wise contact during a short time interval $\Delta t$ is $p_{\Delta t}=\beta \Delta t$. The total infection probability of a susceptible individual connected to $n_\text{inf}$ neighbors during the small time interval $\Delta t$ for such a simple contagion can be calculated to:

\begin{equation}\label{eq:simple}
P_{{sc},\Delta t}(n_\text{inf}) = 1-(1-p_{\Delta t})^{n_\text{inf}} \ .
\end{equation}

Infected individuals transition to the recovered state with a finite, constant recovery rate $\gamma$. Thus, the average infection duration is $\tau_\text{inf}=\gamma^{-1}$. For simplicity, we assume that the recovered state is an absorbing state, i.e. once recovered an agent does not change its state anymore. Starting from an initial state of mostly susceptible agents and a small number of infected the epidemic spread will terminate once there are no more infected agents in the network. 
\subsubsection{Complex contagion model} 
The complex contagion model is analogous to the simple contagion model described above, with exception of the infection probability. Here, we assume a complex contagion process with an overall infection rate of a susceptible individual $\beta_{cc}$ given by a sigmoidal function $\mathcal{S}(r_\text{inf})$ of the fraction of its infected network neighbors $r_\text{inf}=n_\text{inf}/d^\text{in}$ with $d^\text{in}$ being the in-degree of the susceptible individual:
\begin{equation}\label{eq:betac}
\beta_{cc}(r_\text{inf})=\beta_\text{max} \mathcal{S}(r_\text{inf},r_0) =\beta_{\text{max}}\frac{1+ \tanh\big(\sigma\left(r_\text{inf}-r_0\right)\big) }{2}   
\end{equation}
Here, $\beta_\text{max}$ sets the maximal possible infection rate, $\sigma$ controls the steepness of the sigmoidal function, whereas $r_0$ sets the inflection point of the sigmoid with $\mathcal{S}(r_0)=1/2$. For large $\sigma\gg 1$, Eq. \eqref{eq:betac} yields a sharp, step-like function with $\beta_{cc}(0)\approx 0$ and $\beta_{cc}(1)\approx \beta_\text{max}$. For $\sigma\approx 1$, $\beta_{cc}(r_\text{inf})$ is approximately linear with $\beta_{cc}(0)=\beta_\text{max}[1-\tanh(\sigma r_0)]/2$ and $\beta_{cc}(1)=\beta_\text{max}[1+\tanh(\sigma(1-r_0)]/2$. For $\sigma=0$, the infection rate becomes independent on $r_\text{inf}$ with $\beta_c=\beta_\text{max}/2$. The overall probability of an susceptible agent to get infected in a short time interval $\Delta t$ is thus simply:
\begin{equation}
    P_{{cc},\Delta t}(r_\text{inf}) = \beta_{cc}(r_\text{inf})\Delta t \ .
\end{equation}
As the infection probability depends on the fraction of infected individuals and not on the absolute number of infected neighbors, this model describes a fractional, complex contagion process. 

\section{Results}
In a first step we study how different network properties such as average in-degree, clustering coefficient,  shortest path length and the distribution of link length depend on network density. In the case of visual networks we find that all these measures exhibit an interesting non-monotonic relationship with density. The choice of density as an independent parameter is reasonable when studying social behavior because while animals might not be able to change individual perceptual thresholds they can usually adapt their distance to neighboring individuals. For example fish schools have been shown to move closer together under the threat of predation \cite{SosTwoBak19}. 

Given that the average in-degree is modified by network density and the in-degree in turn affects both the clustering coefficient and the average shortest path length we then explicitly investigate the relationship between in-degree and clustering coefficient as well as shortest path length. Here we still find that visual networks differ qualitatively from both metric and topological networks. 

In a second step we study how these static properties affect information propagation by comparing the speed and reach of simple and complex contagion processes through visual and metric networks. Finally, we use the contagion processes to show examples of anisotropy in  contagious spreading that can be observed in visual networks for certain combinations of density and sensory limits and are absent in metric networks.
\begin{figure*}[htpb]
    \centering
    \includegraphics[width=0.9\linewidth]{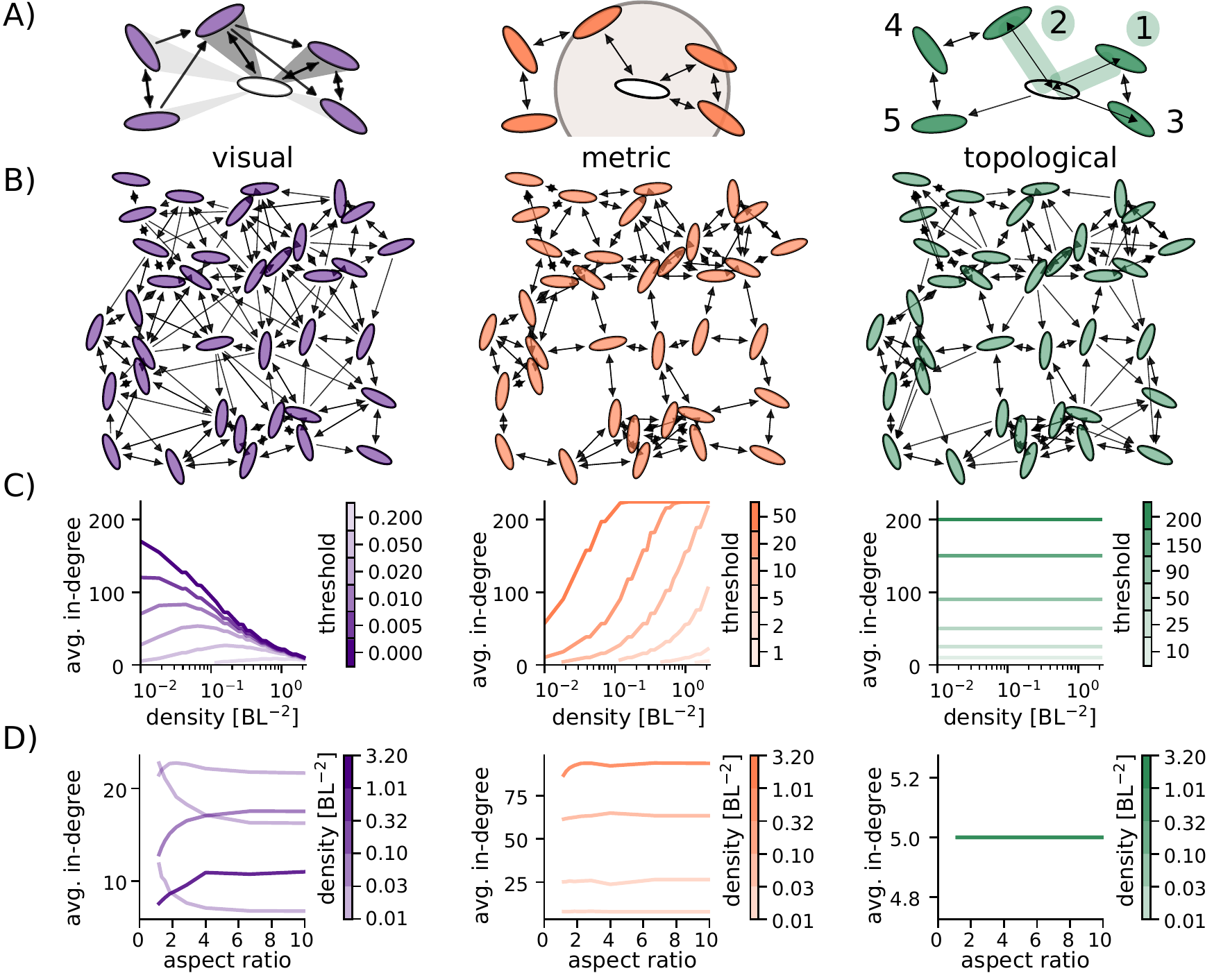}
    \caption{\textbf{The different network types:} \textbf{(A)} illustration of rules determining incoming network links of the central individual (white). Left: visual angles needs to be larger than a threshold value, $\theta_\text{visual}=0.29$, (dark grey shaded areas indicate visual angles that fulfill this requirement, light grey ones do not). Center: metric distance between individuals needs to be below a certain threshold (indicated by a circle). Right: incoming links are coming from fixed number of nearest neighbors (here 2) for the topological model. Numbers indicate closeness-rank. \textbf{(B)} Examples of the different networks for $g=1.4, \eta_{pos}=0.5, N=36, \kappa=0.5, w=0.3$ using the same positions, orientations and total number of links (thresholds: $\theta_\text{visual}=0.254$, $\theta_\text{metric}=2.1035$, $\theta_\text{topo}=5$, yielding 180 directed links).
    \textbf{(C)} Average in-degree as a function of density for different threshold values indicated by line color. The average in-degree of the different network types show a very distinct dependence on density. \textbf{(D)} Influence of ellipses' aspect ratio on avg. in-degree for different density regimes indicated by line color for fixed thresholds, $\theta_\text{visual}=0.05$, $\theta_\text{metric}=8$, $\theta_\text{topo}=5$. At low density circles (aspect 1) are visible over larger distances leading to an increase in average in-degree. At high densities, elongated ellipses (high aspect ratio) cause fewer occlusions which in turn increases the average in-degree.}
    \label{fig:networks}
\end{figure*}

\subsection{Density dependence of network properties}
\label{subsec:static_vs_density}
\subsubsection{In-degree}
Figure \ref{fig:networks}C shows the average in-degree as a function of spatial density of individuals for all three network types and various sensory thresholds. In the case of topological networks the in-degree is not affected by density due to the constraint that an individual can only interact with a fixed number of neighbors independent of their distance. For metric networks average in-degree increases with density. This is explained by the networks' construction where every individual is connect to all other individuals within a fixed range and naturally, the number of individuals within this fixed interaction radius increases with density. 

Visual networks, on the other hand, exhibit a different relationship between density and average in-degree: depending on the visual threshold the average in-degree either monotonously decreases with density (very small  visual  thresholds) or exhibits  a  maximum  at intermediate densities (higher visual thresholds), the exact position of which depends on the value of the visual threshold. The decrease at high densities is due to occlusions in the visual field that become more prominent at high packing fractions and constrain visual interactions. The decrease at low densities is the result of the non-zero visual threshold. Individuals need to occupy a certain angular area in the visual field of a focal individual before they become connected and this requires them to be within a certain distance to it. This behavior is similar to that of the metric network with the maximal interaction radius being determined by the visual threshold and the projected body size of the individual. Since this projected body size heavily depends on the ellipses' aspect ratio, Figure \ref{fig:networks}D takes a closer look at the variation of average in-degree with aspect ratio, which is $\frac{10}{3}\approx3.3$ in Figure \ref{fig:networks}A to C and throughout the rest of this paper. 

In topological and metric networks the average in-degree does generally not depend on the ellipse shape. An exception is the case of metric networks at high density, where for elongated ellipses (high aspect ratio) the average in-degree increases because ellipses can move closer together with a high aspect ratio and thus slightly increases the average in-degree by increasing the number of individuals that can fit within a fixed interaction radius. For visual networks one observes a strong dependence of in-degree on aspect ratio at all ranges of density. For high densities elongated ellipses (high aspect ratio) lead to less occlusions and thus a higher average in-degree. For low densities, circles (aspect ratio 1) remain visible best at large distances, leading to a higher average in-degree for small aspect ratio. At intermediate densities these two opposing effects can lead to a maximum of the average in-degree at intermediate aspect ratios.

\subsubsection{Other network measures}
We study the effect of network density on the average clustering coefficient, the average shortest path length and the distribution of relative link lengths, measures that have proven useful in characterizing types of spatial networks \cite{Barthelemy2018transitions} and thus allow us to contextualize visual networks in the broader landscape of spatial networks. 

\begin{figure}[htpb]
    \centering
    \includegraphics[width=0.9\linewidth]{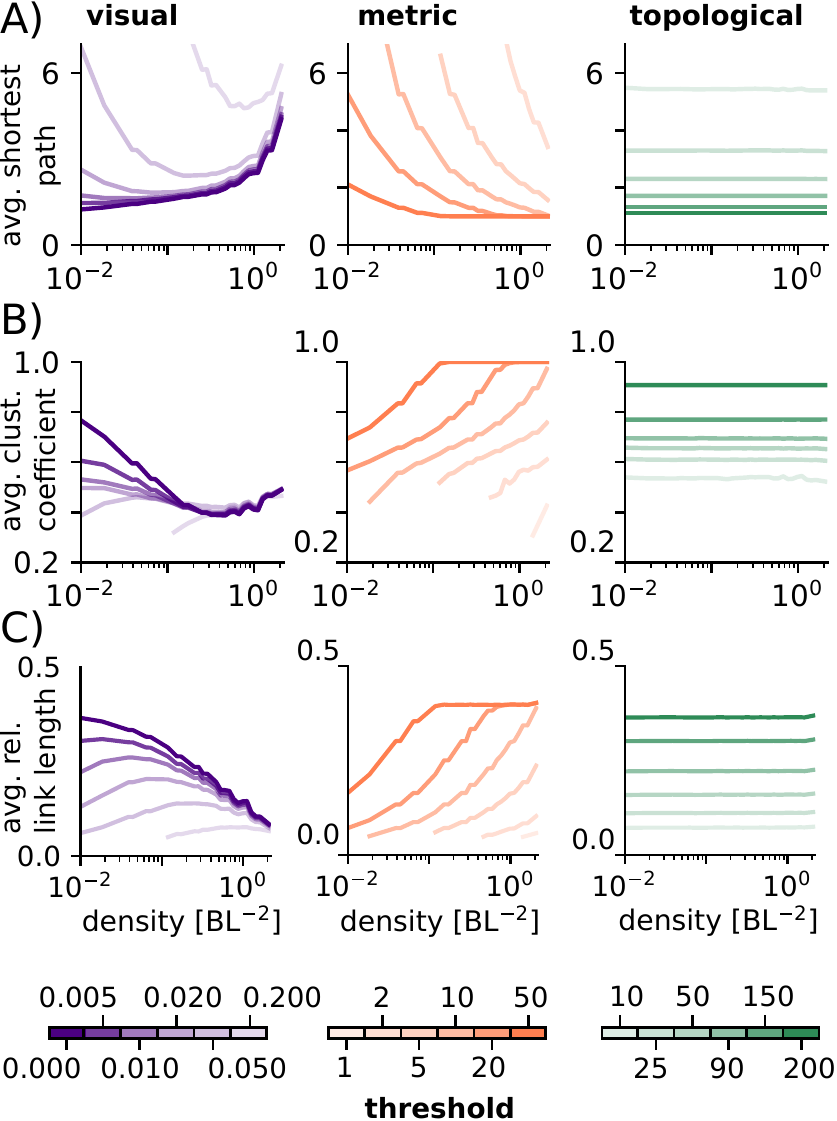}
    \caption{\textbf{Density dependence of network measures: (A)} Average shortest path length as a function of density for all three network types. In visual networks the shortest path length assumes a threshold dependent minimum. In metric networks shortest path length decreases with density. In topological networks no density dependence is observed. \textbf{(B)} Average clustering coefficient as a function of density for all three network types. In visual networks the average clustering coefficient assumes a minimum at intermediate to high densities. The average clustering coefficient increases with density in metric networks and shows no density dependence in topological networks. \textbf{(C)} Average relative link length as a function of density. Visual networks exhibit a threshold dependent maximum while for metric networks link length increases with density.
    Again, topological networks show no dependence on density.}
    \label{fig:networkmeasures_vs_density}
\end{figure}
Figure \ref{fig:networkmeasures_vs_density}A summarizes the effect of spatial density of individuals on the average shortest path length for all three network types and the same sets of thresholds as in Figure \ref{fig:networks}. For topological networks the length of the shortest path is unaffected by network density. In metric networks the average length of the shortest path decreases monotonically with increasing density, which can be explained by the simultaneous increase in in-degree (see Figure \ref{fig:networks}C). In visual networks the average shortest path can exhibit a minimum, the position of which depends on the visual threshold and roughly matches with the maximum in the average in-degree (Fig. \ref{fig:networks}C).

Figure \ref{fig:networkmeasures_vs_density}B shows the effect of density on the average clustering coefficient. For topological networks the clustering coefficient depends only on the threshold, $\theta_{topo}$, and shows no density dependence. 
For metric networks the clustering coefficient increases monotonically with density. This can again be explained by network construction: since every node is connected to all nodes within a radius prescribed by the threshold, two neighbors of the same node are also likely to be close to each other and thus share a connection. 
In the case of visual networks, the clustering coefficient exhibits a non-monotonic relationship with density. In particular, for some threshold values we find a maximum at low densities followed by a (threshold independent) minimum at intermediate density values. 

Average relative link length, depicted in Figure \ref{fig:networkmeasures_vs_density}C, varies with density similarly to the average in-degree, shown in Figure \ref{fig:networks}C. 
 Qualitative differences between the link length distributions of the different network types, which are not adequately captured by the average, can be seen in an example in Figure \ref{fig:networkmeasures_vs_indegree}C and the appendix. 
 
\subsection{Quantitative comparison of network types}
\label{subsec:static_vs_degree}
\begin{figure*}[htpb]
    \centering
    \includegraphics[width=.9\linewidth]{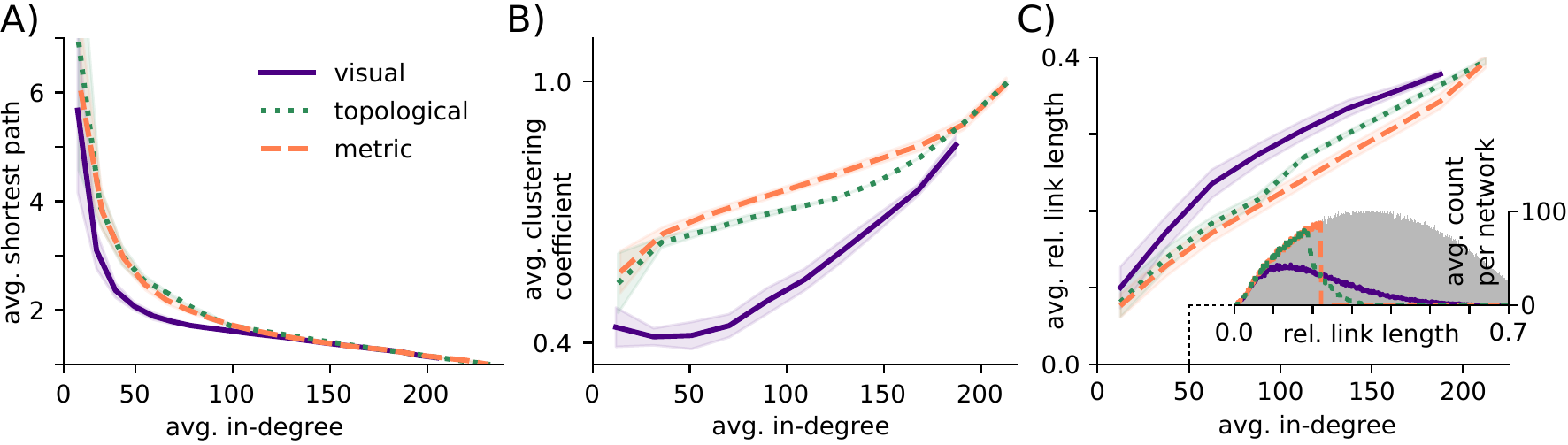}
    \caption{\textbf{Degree dependence of network measures.} All Figures were created using networks of $N=225$ averaged over all densities and thresholds. \textbf{(A)} Average shortest path length as a function of average in-degree. At low in-degrees visual networks exhibit shorter shortest-path lengths than both metric and topological networks. \textbf{(B)} Average clustering coefficient as a function of average in-degree. Consistent with the results in Figure \ref{fig:networkmeasures_vs_density} visual network display substantially lower clustering across a wide range of in-degrees than both metric and topological networks. \textbf{(C)} Average relative link length as a function of average in-degree. For intermediate in-degree, visual networks exhibit considerable longer links than both metric and topological networks. Inset: Example of the average distribution of relative link length for networks of average in-degree 50.  The grey shaded area indicates the link length distribution for a fully connected network. Parameters: $g=2.0, \theta_\text{vis}=0.00505, \theta_\text{metric}=9.2944, \theta_\text{topo}=50$, averaged over 40 networks.}
    \label{fig:networkmeasures_vs_indegree}
\end{figure*}
While we have observed a variety of changes in networks measures with density, we have also found that they can to a large degree be explained by changes in the average in-degree. It is thus important to discern how much of this difference between network types persists when the average in-degree is kept fixed. The results, shown in Figure \ref{fig:networkmeasures_vs_indegree}, represent averages over all densities.
 
 Figure \ref{fig:networkmeasures_vs_indegree}A depicts the average shortest path, which decreases as networks become more densely connected with little difference between metric and topological networks. In visual networks, however, shortest paths tend to be shorter, especially at low to intermediate in-degrees. This decreased average shortest path length can most likely be explained by the presence of long links allowing shortcuts between spatially distant nodes (see Figure \ref{fig:networkmeasures_vs_indegree}C). 
 
Figure \ref{fig:networkmeasures_vs_indegree}B shows that for all three network types the average clustering coefficient increases with the number of incoming links per node. Again, there is little difference between metric and topological networks, whereas over a wide range of in-degrees the clustering of visual networks is substantially lower. This phenomenon can be partly explained by the presence of long-range connections breaking local clusters  \cite[c.f.][]{Brockmann1337}. Additionally, local clustering structure is disrupted by the visual blocking of neighbors on either side of an individual. At very high average in-degree, when the networks become essentially all-to-all connected, clustering approaches one. Visual networks however, do not reach such high degrees due to occlusions.

Finally, Figure \ref{fig:networkmeasures_vs_indegree}C, shows the average relative link length of the different network types. While visual networks have a slightly higher average relative link length, the difference between the networks becomes more apparent when looking at the full distributions, shown as an inset for an average in-degree of 50. For comparison the distribution of the fully connected network is added in grey. In metric networks all links up to the threshold value are realized, where the distribution shows a sharp cutoff. For our spatial distribution of relatively homogeneous density, topological networks include all links up to a certain length and then show a fast decay in the distribution. The distribution of visual networks shows a much slower decay for higher link length, confirming the existence of substantially longer links in visual networks and underpinning our attribution of differences in the other two network measures to a difference in link length and shortcuts via a few very long links.

 \subsection{Density dependence of contagion processes}
 \label{subsec:dynamic_vs_density}
 In section \ref{subsec:static_vs_density} we showed that the density dependence of the visual networks' structural properties is very distinct from that of the other two network types. 
 In order to understand their implications for dynamic processes, we compare the evolution of simple and complex contagion dynamics on visual to that on metric networks. We omit topological networks because of the independence of their topological properties from density and their similarity to metric networks in their dependence on average in-degree (see Figures \ref{fig:networkmeasures_vs_density}, \ref{fig:networkmeasures_vs_indegree}).

\begin{figure*}[htpb]
    \centering
    \includegraphics[width=.9\linewidth]{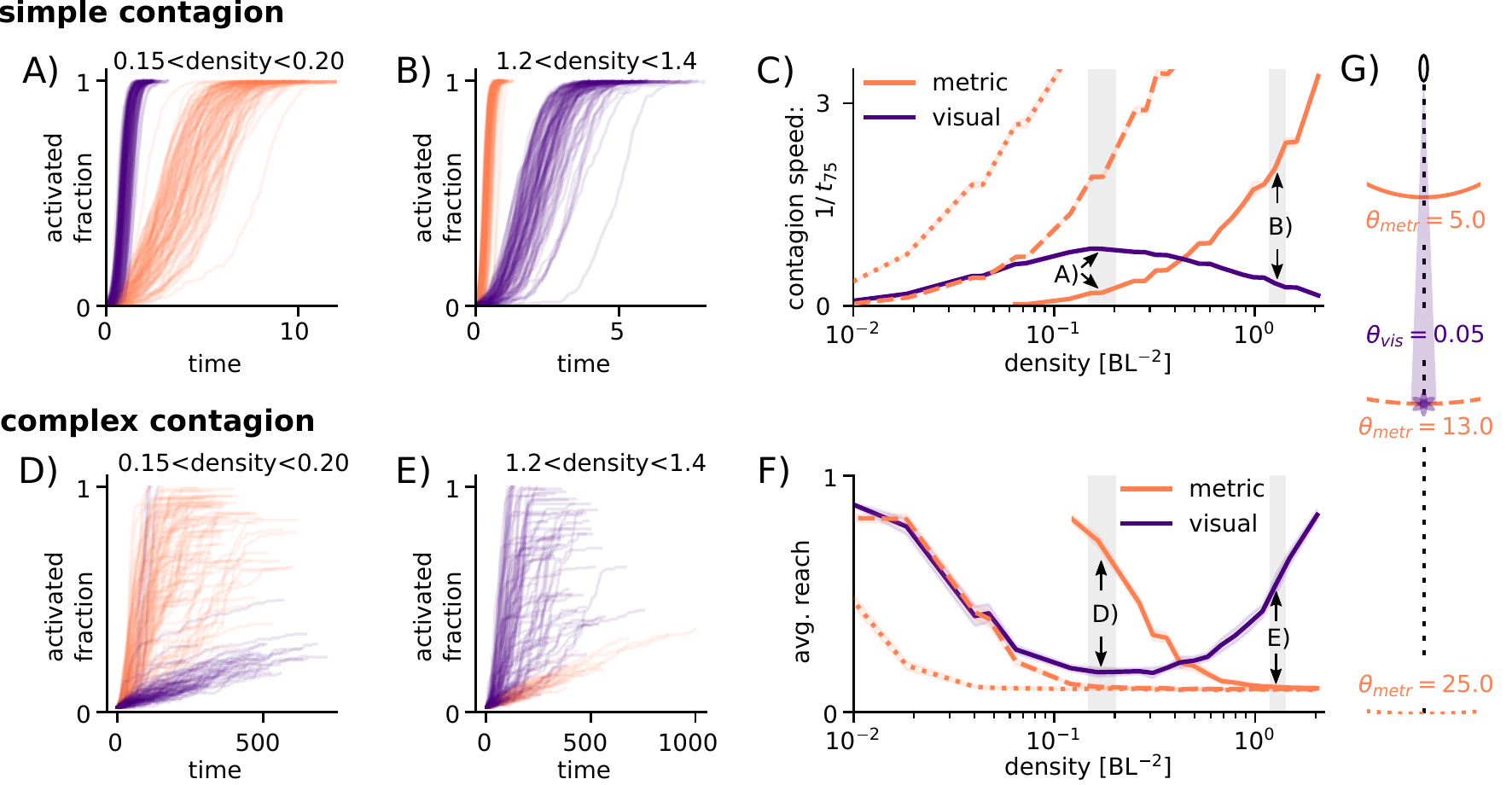}
    \caption{\textbf{Simple and complex contagion on visual and metric networks.}\textbf{(A)} Example trajectories for a simple contagion process at low densities for visual (purple) and metric (orange) networks showing faster spread in metric networks. \textbf{(B)} same as (A) but for high densities showing faster spread for visual networks. \textbf{(C)} Comparison of visual and metric network w.r.t the speed of infection via simple contagion. In metric networks the spead of spread increases monotonically with density. In visual networks the contagion speed assumes a maximum at intermediate densities. Line-styles correspond to different thresholds of the metric network as illustrated in G). Shaded grey areas correspond to the density ranges in (A) and (B) respectively.
    \textbf{(D)} Example trajectories for a complex contagion process at low densities showing faster spread in metric networks. Note, that most trials do not reach full activation in either network type. \textbf{(E)} Same as (D) for high densities, showing faster spread in visual networks. Again, most trials do not reach full activation. \textbf{(F)} Comparison of average reach (i.e. final fraction of activated individuals) as a function of density for both network types. In metric networks average reach decreases with density, while visual networks show a minimum at intermediate densities. As in (C) line-styles correspond to different metric thresholds and shaded areas correspond to the examples in (D) and (E) respectively. Parameters: $N=225$, $\Phi \in [0, 0.3]$, $p_{rec}=0.03$, $\beta=0.3$, $r_0 = 0.35$, $\sigma=10$. Lines in (C) and (F) are averages over 100 networks and 100 runs per network, starting the contagion process with one (simple contagion) or five (complex contagion) randomly chosen individuals.}
    \label{fig:simple_cont}
\end{figure*}
\subsubsection{Simple Contagion}
For a connected network and a sufficiently low recovery rate a simple contagion process will always spread through the entire network activating all nodes eventually. In order to compare the two network types we study the speed of the spread measured by the inverse of the time it takes for the activation to spread from a single individual to 75 \% of the network, $1/t_{75}$. The probability of a node becoming infected is proportional to its number of infected neighbors (see eq. (\ref{eq:simple})). Therefore we can expect the infection to spread faster in networks with high average in-degree. 

Figures \ref{fig:simple_cont}A and B show examples of the time course of infection for low and high densities respectively (for parameter sets refer to the figure caption). For the parameters used, at low densities (A) the infection spreads faster on visual networks, at high densities (B) the metric networks have a speed advantage. The solid lines in Figure \ref{fig:simple_cont}C summarize the effect of density on speed for both network types and confirms the observation that for lower densities a simple contagion process will spread substantially faster through a visual than through a metric network while at higher densities this effect is reversed. The speed maximum of visual networks in the low to mid density regime correlates nicely with the shortest path length as shown in Figure \ref{fig:networkmeasures_vs_density} (for the threshold value 0.05) and can be explained by the presence of long-range connections.

However, the above examples are for one specific choice of metric and visual thresholds, $\theta_\text{metric}=5\,\text{BL}$ and $\theta_\text{visual}=0.05\,\text{rad}$. As illustrated in Figure \ref{fig:simple_cont}G, this choice allows visual links of (on average) up to $13\,\text{BL}$ and thus longer-ranged visual than metric interactions (5 BL). For a full picture of the possible quantitative differences between the two network types, we include two more parameter values of the metric network. For $\theta_\text{metric}=13\,\text{BL}$ the maximal link length of metric and visual networks is similar (dashed line) and the two networks have a comparable contagion speed at low density, only diverging at higher densities when occlusions in the visual network lead to a decrease in average in-degree compared to the metric networks. For $\theta_\text{metric}=25\,\text{BL}$ metric interactions can be longer than visual ones (dotted line) and the simple contagion process spreads faster on the metric networks at all densities because of their higher average in-degree.

Considering metric networks as a simple model of acoustic interactions, the different choices of thresholds could describe animals that can see further/equally far/shorter than they can hear which will strongly depend on the animals physiology but also on the properties of the surrounding medium. While in those cases where metric interactions can be longer than visual ones acoustic interactions can be understood to provide a faster transfer of information across the group at all densities (dotted line), for the opposite case (solid line), the fastest mode of information transfer strongly depends on density.
    
\subsubsection{Complex Contagion}
In the complex contagion model considered here, the probability of getting infected is a sigmoidal function of the fraction of infected neighbors. Thus individuals only have a high infection probability if their fraction of infected neighbors, $r_\text{inf}$, exceeds the threshold $r_0$. 
A fractional contagion process is not guaranteed to spread through the entire network. If no node has a fraction of infected neighbors $\geq r_0$, the remaining infection rates of the individuals may drop far below the recovery rate and the process can come to a halt. To account for the possibility of incomplete spread we compare the network types w.r.t. the fraction of the network that gets infected before the process stops, instead of the contagion speed for the complex contagion. We refer to this as the \textit{reach} of the contagion.

Where in the case of simple contagion having a high in-degree increases a node's rate of infection, because it increases its probability to have infected neighbors, the contrary is true for fractional contagion in the case of a low overall prevalence of the 'infected' state. Assuming a fixed number of infected neighbors (as we use to initiate the processes in Figure \ref{fig:simple_cont}D to F), an increase in in-degree will only lower a node's fraction of infected neighbors and thus decrease infection probability. 

Figure \ref{fig:simple_cont}D and E show the time course of infection for the same density regimes as in A and B. Compared to the simple contagion process the roles of the networks appear reversed: at low densities the fractional contagion process spreads faster and further in metric than in visual networks, while at high densities visual networks are faster and become infected to a larger fraction. It is also clear, that most networks do not get fully infected for this choice of parameters. 
Figure \ref{fig:simple_cont}F compares the fraction of infected nodes in both network types as a function of density for the same set of thresholds as in Figure \ref{fig:simple_cont}C. The effect of density is clearly reversed between the complex and the simple contagion process which can be attributed to the opposing effect of an increase in in-degree on the infection probability of a node as discussed above. 

Coming back to the interpretation of the two network types as based on two different senses (vision and hearing), we can now see that in addition to the group density the optimal mode of transmission (acoustic or visual) does also depend on the type of contagion process. A combination of both interaction types may allow robust communication independent of group density.

Another notable feature that distinguishes visual from metric networks here is their consistency and robustness over a wide range of densities. While contagion speed for metric networks quickly increases with density (for all threshold values), visual networks provide comparable speeds at high and low densities. Similarly, for each metric threshold there is an upper density limit for the transmission of a complex contagion, but we can find visual thresholds that allow a complex contagion to (partially) pass at all densities.

\subsection{Quantitative comparison of contagion processes on visual and metric networks}
\begin{figure*}[htpb]
    \centering
    \includegraphics[width=.9\linewidth]{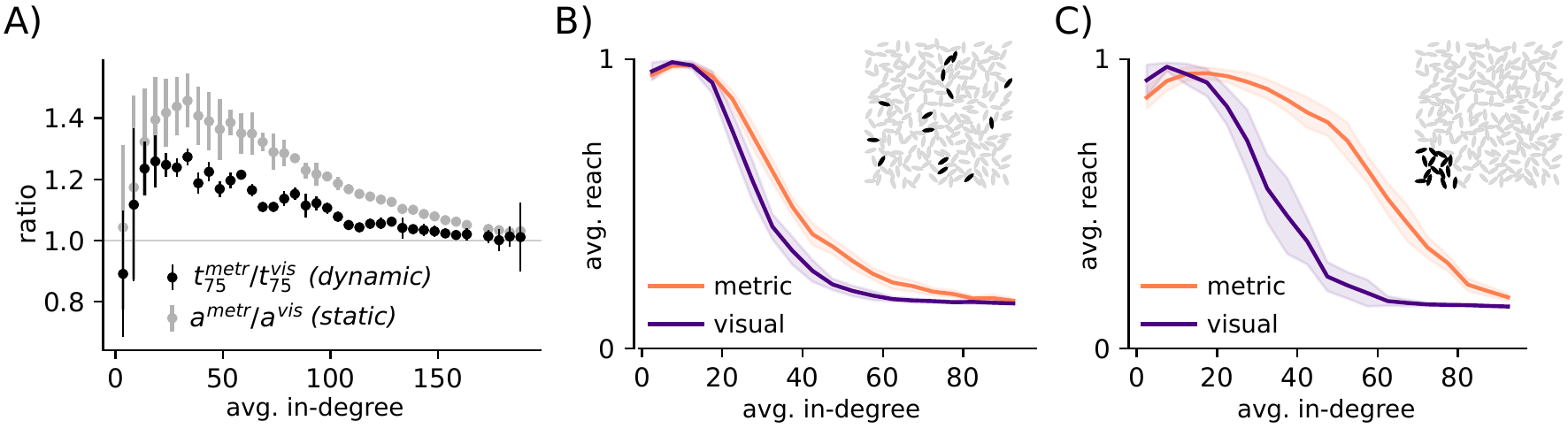}
    \caption{\textbf{Structural differences influence dynamics on networks:} \textbf{A)} The ratio of contagion speeds (time to 75\% infected or recovered, $t^{metr}_{75}/t^{vis}_{75}$) shows a similar dependence on average in-degree as the ratio of average shortest path lengths ($a^{metr}/a^{vis}$, grey). Dots represent averages and errorbars one standard deviation over 100 networks (static) or 100 runs on 100 networks (dynamic). \textbf{(B) and (C)} Effect of clustering coefficient on fractional contagion. (B) Initial activation by 15 randomly selected individuals (see black ellipses in inset for an example). (C) Initial activation by 15 individuals in one corner of the group (inset). Lines represent averages over 100 networks with 100 runs of the fractional contagion dynamic each, shaded areas represent one standard deviation of the networks' average reach. A higher clustering coefficient (as in metric networks, compare Figure \ref{fig:networkmeasures_vs_indegree}) is beneficial for a spatially clustered initial activation, (C).  Parameters: $N=225, \Phi\in[0.0,0.3], p_{rec}=0.01, \beta=0.3, r_0=0.4, \sigma=10$}
    \label{fig:dynamics_vs_degree}
\end{figure*}

 In section \ref{subsec:static_vs_degree} a degree-controlled comparison of the different network types revealed that for intermediate average in-degrees visual networks on average have a lower shortest path length, a lower average clustering coefficient and links that span larger distances than their metric and topological counterparts (Figure \ref{fig:networkmeasures_vs_indegree}). 
In order to illustrate the effect that these differences can have on dynamic processes, we again compare the evolution of simple and complex contagion dynamics on visual to that on metric networks for an exemplary set of parameters, see caption of Figure \ref{fig:dynamics_vs_degree}.

Figure \ref{fig:dynamics_vs_degree}A depicts the spreading speed of a simple contagion on visual networks in units of the spreading speed on degree-matched metric networks, $t_\text{75}^\text{metr}/t_\text{75}^\text{vis}$, as a function of average in-degree (black dots). Spreading on visual networks is faster (indicated by values larger than 1) for a wide range of intermediate average in-degrees. The ratio of average shortest path lengths, $a^\text{metric}/a^\text{visual}$, (grey dots) mirrors the qualitative shape of the speed ratio indicating that the increased speed of the simple contagion process on visual networks can be attributed to their shorter average shortest paths (compare Figure \ref{fig:networkmeasures_vs_indegree}A). 

Figure \ref{fig:dynamics_vs_degree}B and C illustrate the effect of the lower clustering coefficient of visual networks on the reach of the complex fractional contagion. As already discussed in section \ref{subsec:dynamic_vs_density} in the context of Figure \ref{fig:simple_cont}F, as long as the number of infected individuals in the network is low and infections are randomly distributed (illustrated in the inset of Figure \ref{fig:dynamics_vs_degree}B), each additional link is most likely decreasing the focal node's infection probability in the fractional contagion process (because it is more likely to a susceptible than to an infected individual). This is the case in Figure \ref{fig:dynamics_vs_degree}B for both network types and explains the decrease of the average reach with increasing average in-degree. 
A spatial clustering of infections (as in the initial conditions of Figure \ref{fig:dynamics_vs_degree}C, illustrated in the inset) increases the fraction of infected individuals in the neighborhood of nodes with a high local clustering and close to the 'wave front' of the infection (the border between susceptible and infected individuals). Therefore, the average reach of the fractional contagion remains substantially larger on metric than on visual networks with increasing average in-degree for a spatially correlated initial activation (Figure \ref{fig:dynamics_vs_degree}C).

 Put differently, visual networks have a longer average link length for a comparable average in-degree, which results in a lower clustering and in nodes receiving inputs from many, possibly far away and not spatially correlated neighbors. This makes it difficult for any node to reach the required fraction of infected neighbors in case of a very spatially confined spreading (i.e. a single wave front passing throught the group). The long range connections lead to a diffusion of information and an overall decrease in the local fraction of infections. Thus the same mechanisms that have proven helpful in the case of simple contagion (i.e. the long links, providing short cuts), hinder the spread of information for complex contagion. The hampering effect of long range connections has also been described in the sociological literature where its is known as the "weakness of long ties" \cite{centola2007complex}.
 
\subsection{Polarization}
\begin{figure*}
    \centering
    \includegraphics[width=.9\linewidth]{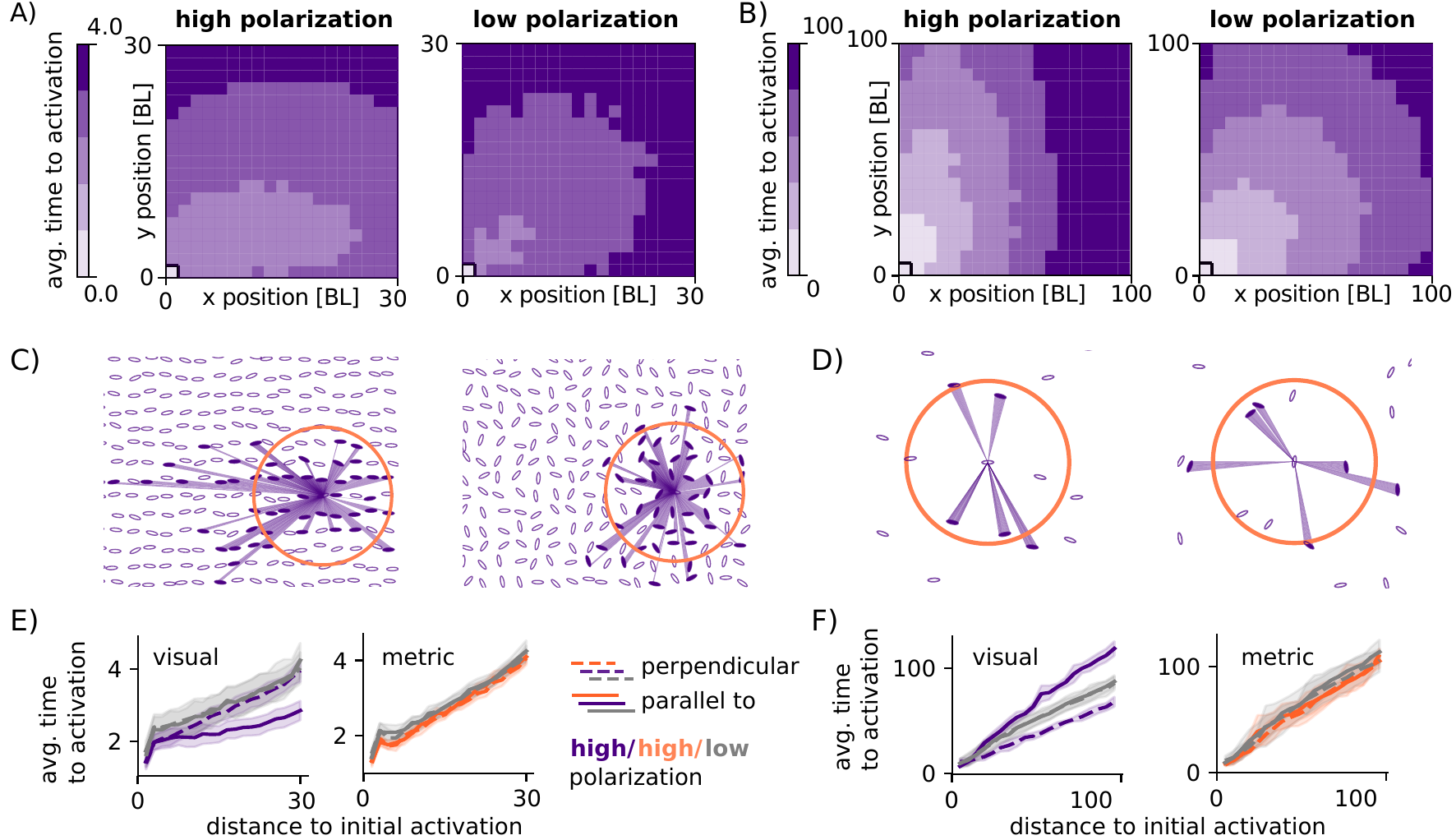}
    \caption{\textbf{Anisotropic spreading of simple contagion for high group polarization: (A)} Propagation time of a simple contagion process initiated at $t=0$ by one individual in the bottom left corner of a 20x20 grid with $g=1.5\,\text{BL}$ and high (left, $\Phi\in[0.8,1.0]$) and low (right,$\Phi\in[0.0,0.2]$) polarization along the x-axis. Average time to activation (time, $t>0$, at which an individual at position (x,y) is infected) is obtained as an average over 30 visual networks with 50 runs each and $\theta_\text{visual}=0.01$. \textbf{(B)} Like (A), but for $g=5.0$ and $\theta_\text{visual}=0.1$. \textbf{(C)} and \textbf{(D)} Illustrations of the spatial configurations and visual fields used in the contruction of networks underlying the results of (A) and (B) respectively. Visual angles $\alpha_{ij}>\theta_\text{visual}$ are shown as shaded purple areas for one exemplary focal individual. Orange circles indicate the metric thresholds used in the construction of (E), $\theta_\text{metric}=6$, and (F) $\theta_\text{metric}=7$. \textbf{(E)} Average time to activation averaged over 3 outer rows in x (solid lines) and y (dashed lines) direction for high (purple, orange) and low (grey) polarization. The contagion spreads faster parallel than perpendicular to the polarization for the visual networks. For metric networks (right) polarization has no effect on activation time. Parameters as in (A) and (C). \textbf{(F)} As in (E) but with parameters as in (B) and (D). For this combination of density and visual threshold the contagion spreads faster perpendicular to the polarization. Metric thresholds were chosen to yield similar contagion speeds as low polarization visual networks. Parameters: $N=400, p_{rec}=0.03, \beta=0.1$.}
    \label{fig:polarization}
\end{figure*}
While so far we have studied the effect of density and degree on the structural properties of and contagion processes on the different network types, we have not considered the influence of the group‘s polarization which we have kept between 0 and 0.3. While under natural conditions polarization may correlate with density and thus indirectly influence network structure, for our grid-like positioning of ellipses, metric and topological network links are not influenced by the orientation of individuals because of their sole dependence on the Euclidean distance between individuals. For visual networks, however, the orientation of an ellipse may crucially influence its visibility and thus its social interactions. A high polarization may therefore lead to an anisotropy in spreading only on visual networks, i.e. a difference in speed of propagation in direction of the polarization to that perpendicular to it. Figure \ref{fig:polarization} illustrates this effect with two examples of  a combination of group density and visual threshold that leads to an anisotropy at high polarization, which is notable in the speed of a simple contagion process on visual networks. 

Figure \ref{fig:polarization}A and B depict the average propagation of a simple contagion process through a group of 400 individuals in a square 20x20 grid with average distance of $1.5\,\text{BL}$ and $5.0\,\text{BL}$ respectively between grid neighbors ($g=1.5$ and $g=5.0$). The contagion process is initiated by one individual in the lower left corner at $t=0$ for high ($\Phi\in[0.0,0.2]$, left) and low ($\Phi\in[0.8,1.0]$, right) polarization along the x-axis. The average time it takes the infection to spread from the initial position at (0,0) to any other point in the group (given by an x and y position in body length) is indicated by color intensity (averages over 30 networks with 50 runs each).  For low polarization the contagion process spreads evenly in all directions (circular equitemporal regions) while for high polarization spreading speed is enhanced either along (A) or perpendicular (B) to the direction of polarization (elliptic equitemporal regions).

Panels C and D show examples of the underlying spatial group configurations. They include examples of the visual field of one focal individual. Angular areas that exceed the visual threshold used for the construction of networks for A and B ($\theta_\text{visual}=0.01\,\text{BL}$ in A,C,E and $\theta_\text{visual}=0.1\,\text{BL}$ in B, D, F), are shown as shaded purple areas and corresponding visual network neighbors are filled in purple. For the low visual threshold and high density used in C, at high polarization links can span longer distances and are more numerous in direction of the polarization as occlusions are less prominent in this direction because of the smaller visual angles. For the high visual threshold and low density used in D, at high polarization links are more likely to exist perpendicular to the polarization because of the larger angular areas of ellipses when seen from the side. These unevenly distributed links increase spreading speed in the direction of higher link density and length.
In Panel E and F we summarize the above results by averaging over the outer three rows of individuals in x (solid lines) and y (dashed lines) direction for high (purple) and low (grey) polarization. For low polarization perpendicular and parallel propagation times are similar, while for high polarization propagation time is reduced either parallel (E) or perpendicular (F) to the polarization. Metric networks with a threshold of $\theta_\text{metric}=6\,\text{BL}$ (illustrated by orange circle in C) and $\theta_\text{metric}=7,\text{BL}$ (D) show no difference in propagation times for high and low polarization (E,F). Metric threshold values were chosen to yield similar contagion speeds as the visual counterparts.

\section{Discussion}
Animal groups represent examples of spatially-embedded interaction networks, where the spatial density of individuals does not only vary due to external factors but can be actively modulated by the group members based on environmental context \cite{SosTwoBak19}. We have shown, that when describing such systems with highly variable density by considering fixed interaction thresholds, potentially related to sensory or cognitive limitations, the fundamental properties of the resulting interaction network, their qualitative dependence on density and their emergent collective dynamics will crucially depend on the type of the network used. In particular, each network type shows a characteristic qualitative dependence of the average degree on density, which influences the density dependence of other network measures as well as the speed and reach of two classes of contagion processes (Figures \ref{fig:networks}, \ref{fig:networkmeasures_vs_density} and \ref{fig:simple_cont}). We characterized visual networks as a distinct class of interaction networks and highlighted the often neglected dependence of collective behavior on density and network type.

When modelling the collective behavior of animal groups with variable densities and fixed thresholds, intermediate-threshold visual networks display several key advantages. 
In contrast to metric networks, the average degree does not monotonically increase with density and thus avoids unrealistically high values at high densities. In contrast to topological networks they show a non-trivial dependence of the (in-)degree on density, and more crucially the finite visual threshold introduces a maximal interaction distance, which makes their behavior more realistic at low densities. In addition, they exhibit a maximum in the average (in-)degree at intermediate spatial densities of individuals which could explain preferred group densities. 

Our results further demonstrate that for wide ranges of group densities visual networks are characterized by the existence of long-ranged links, absent in metric or topological interaction networks with comparable in-degrees which are accompanied by a lower average shortest path and clustering coefficient (Figure \ref{fig:networkmeasures_vs_indegree}). The role of such long-range links in facilitating simple contagion processes such as information diffusion, has been studied extensively in network science and is known as the "strength of weak ties" or the "small-world" phenomenon \cite{granovetter1973strength,watts1998collective}. Here we observe their influence in an increased spreading speed of the simple contagion on visual networks as compared to metric ones of similar degree (Figure \ref{fig:dynamics_vs_degree}).
On the other hand, visual networks have in general smaller clustering coefficients in comparison to metric and topological networks, which is disadvantageous for the spread of fractional, complex contagion processes, assumed to be involved in spreading of behaviors \cite{Rosenthal2015} and known as the "weakness of long ties" \cite{centola2007complex}. We can observe this effect in our simulations of a complex fractional contagion process on visual and metric networks of the same degree. Here, the contagion process spreads to a larger fraction of the metric than the visual network and this effect becomes even more apparent when the initial activation is spatially correlated (Figure \ref{fig:dynamics_vs_degree}). 

In summary, how fast and how far a behavior or information will spread, is dependent upon the network density, the agent's sensory limits and the type of contagion process. The use of multiple types of interactions (like visual or acoustic) may enable organisms to compensate the shortcomings of one type of sensory interaction and thus enable reliable collective response across a range of densities or sensory limits varying under changing environmental conditions. 

Finally, we show that only visual networks have a strong dependence on the aspect ratio and orientation of individuals. More specifically, the breaking of a group's orientational symmetry due to alignment of individuals induces a symmetry breaking in the interaction network and consequently spatial anisotropy of social interactions. For two exemplary combinations of density and visual threshold we found anisotropy in the spreading of a simple contagion process on visual networks of polarized groups. More detailed studies of this dependence could reveal advantages or disadvantages of different spatial configurations of animal collectives observed in nature, especially when studied together with their visual detection ability \cite{Davidson2021}.

Overall, our work provides compelling arguments for the consideration of visual networks in the study of social behavior. Nevertheless, it marks only a first step towards a full understanding of their role in collective animal behavior. While our results for artificially generated, static networks based on a square grid allow us to systematically study the effect of density, they do not capture the temporal nature and full range of structures observed in animal groups. Luckily, recent improvements in tracking software promise faster and more convenient access to realistic animal network data, including visual networks \cite{walter2021trex}. Using networks constructed from animal tracking would provide information about naturally occurring group sizes, densities and polarization as well as spatio-temporal fluctuations within these measures and thus allow to study their effects within a naturally occurring parameter range. Additional knowledge about the sensory limits specific to the studied organism (including the addition of a visual blind angle) will further improve the interaction networks and may enable a direct comparison between networks based on different senses like sound perception and vision.

A next step could then be to move on to non-static networks and to study the effect of different macroscopic states such as milling or swarming on network structure and spreading dynamics. Such an approach could also take into account the interactions of a group with the environment, asking, for example, how the detection of visual cues and the trade-off between private (external) and social information depends on group structure. It has been shown that predator detection based on sensory limits is dependent on density and group size and varies between different macroscopic states of fish schools \cite{Davidson2021} whereas interactions with the environment may induce density fluctuations. We hypothesize that visual networks are more robust to such spatial perturbations (i.e. a local increase or decrease in density). 

Finally, let us note that, by starting with the most basic approach and considering binary connections (1 or 0 for presence or absence of links), we laid out the fundamental effects of the different networks' dependency on density and threshold. A natural extension to weighted networks in the visual model would be to use relative angular area (angular area divided by the total field of view) as a link weight. 
Such an approach would entail an additional dependence of average link strength on density, which would in turn affect the weighted counterparts of the network measures. 
For the purpose of this paper, i.e. to establish visual networks as fundamentally different from other spatial networks by making direct comparisons between network types, the introduction of link weights would have been impractical as it would have introduced additional and arbitrary dependencies on density by the choice of link weights for each network type. Observed differences between the weighted networks would be heavily influenced by the choice of link weights and could therefore be less clearly attributed to sensory limits and the different rules establishing links. A comparisons between weighted networks will thus be most informative when studying a specific biological system which justifies the choice of link weights. 
Nevertheless, we expect essential features of the visual network, i.e. the existence of an upper bound on the degree at high densities, a lower clustering, the existence of longer links and the anisotropy of visual interaction networks due to orientational symmetry breaking, to also hold for weighted networks.

In conclusion, our work proposes experimentally testable hypothesis, e.g. in the context of behavioral contagion in animal groups \cite{Rosenthal2015,Herbert-Read15}, as well as a theoretical foundation for future investigation on how collective information processing could be dynamically tuned by individual-level behavioral adaptations affecting local density \cite{SosTwoBak19}. We highlighted several important qualities of visual networks, including their unique dependence on density and polarization, which encourages further research in this area. Overall, visual networks provide a promising and necessary addition to the established toolkit for the study of social interactions and collective behaviour and emphasize the need to include system-specific sensory limits.

\section*{Conflict of Interest Statement}
The authors declare that the research was conducted in the absence of any commercial or financial relationships that could be construed as a potential conflict of interest.

\section*{Author Contributions}
WP and PR, conceived the idea and designed the study. WP and CW performed the research. All authors discussed the results and wrote the manuscript. All authors contributed to manuscript revision, read, and approved the submitted version.

\section*{Funding}
W.P. and P.R. were funded by the Deutsche Forschungsgemeinschaft (DFG) (German Research Foundation), Grant RO47766/2-1 . P.R. acknowledges funding by the DFG under Germany’s Excellence Strategy–EXC 2002/1 “Science of Intelligence”–Project 390523135.

\section*{Data Availability Statement}
The code used to generate networks and simulate contagion dynamics on them can be found on GitHub and was archived via Zenodo, \cite{code}.

\begin{acknowledgments}
We acknowledge the support by Luke Longreen with exploratory simulations of visual networks as part of his lab rotation.
\end{acknowledgments}

\bibliography{references}
\newpage
\section*{Appendix}
\appendix
\section{Group density estimation}
\begin{figure}[htpb]
    \centering
    \includegraphics[width=0.6\linewidth]{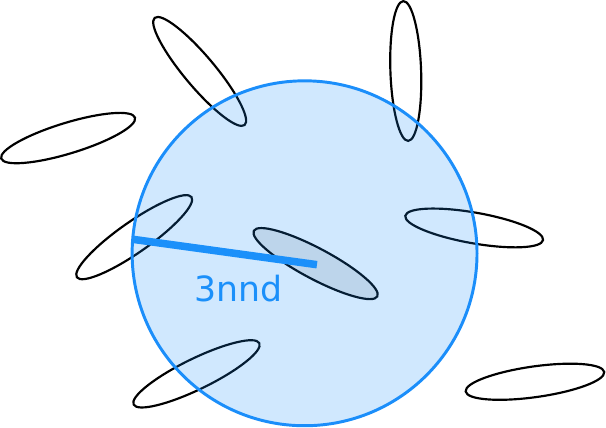}
    \caption{\textbf{Example of the calculation of group density:} Density is calculated from the average third nearest neighbor distance, $r_{3nnd}$. A circle with the diameter of the third nearest neighbor contains 4 individuals in an area of $\pi r_{3nnd}^2$, yielding a local density of $4/(\pi r_{3nnd}^2)$.}
    \label{fig:SI_density_calc}
\end{figure}

For an illustration of the density estimation refer to Figure \ref{fig:SI_density_calc}.
\section{Elimination of overlaps}
\begin{figure}[htpb]
    \centering
    \includegraphics[width=0.8\linewidth]{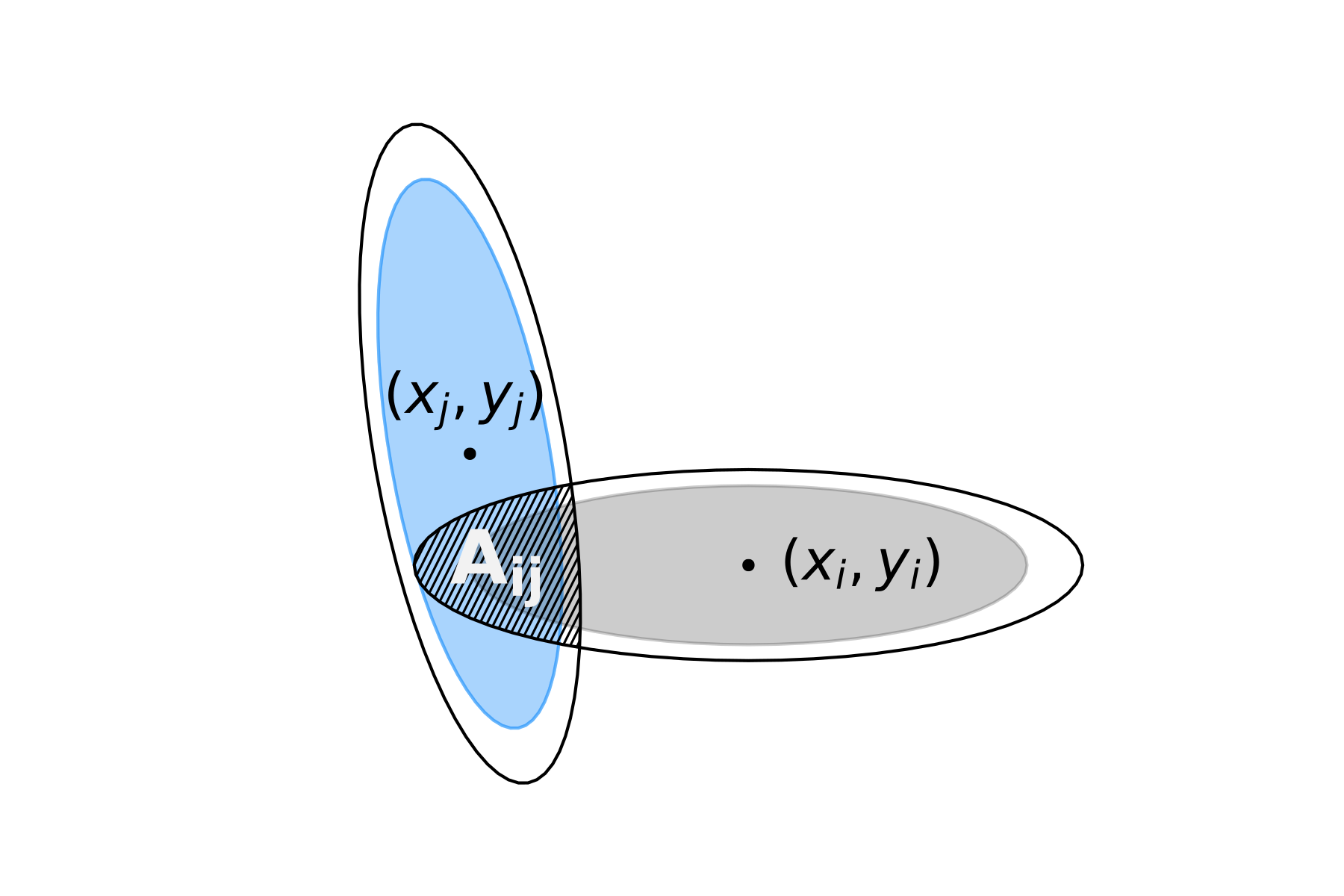}
    \caption{\textbf{Illustration of the ellipse interactions in the particle simulation.} Ellipses repel each other based on their overlap area (hatched). To speed up the simulation and avoid very small forces towards the end we use a larger ellipse (rescaled by a factor 1.1, black lines) to calculate overlap area and stopped the simulation when the original sized ellipses (filled areas) do not overlap anymore.}
    \label{fig:overlap}
\end{figure}

To ensure 2 dimensional groups without overlaps we use a particle simulation based on the code provided within \cite{Palachanis2015}. Ellipses start at their noisy grid positions (see main text) with velocities $v_i(t=0)=0$. These velocities $\vec{v}_i(t)$ evolve based on the positions $\vec{x}_i(t)=(x_i(t),y_i(t))^T$ as follows 
\begin{equation}
    \begin{aligned}
         \frac{d\vec{v}_i(t)}{dt}&=\frac{1}{m_i}\left(-\alpha\vec{v}_i(t)+\sum_{j\neq i}^j\frac{\vec{x}_i-\vec{x}_j}{|\vec{x}_i(t)-\vec{x}_j(t)|}F_{ij}(t)\right)\\
        F_{ij}&= \lambda A_{ij}(t)~.
    \end{aligned}
\end{equation}
Here $A_{ij}(t)$ denotes the area where ellipse $i$ and ellipse $j$ overlap at time $t$, $\alpha$ is the damping parameter, $\lambda$ a constant model parameter and $m_i=m=1$ is the mass of an ellipse (here: $\alpha=0.2$, $\lambda=0.05$). The simulation ends when all overlaps are removed. We used a second larger ellipse (factor 1.1) to determine the repulsion area $A_{ij}$  to speed up the relaxation into a non-overlapping state and stopped when the original size ellipses were no longer overlapping, see Fig. \ref{fig:overlap}.
\section{Analytical calculation of visual field of ellipse}
\begin{figure}
    \centering
    \includegraphics[width=.8\linewidth]{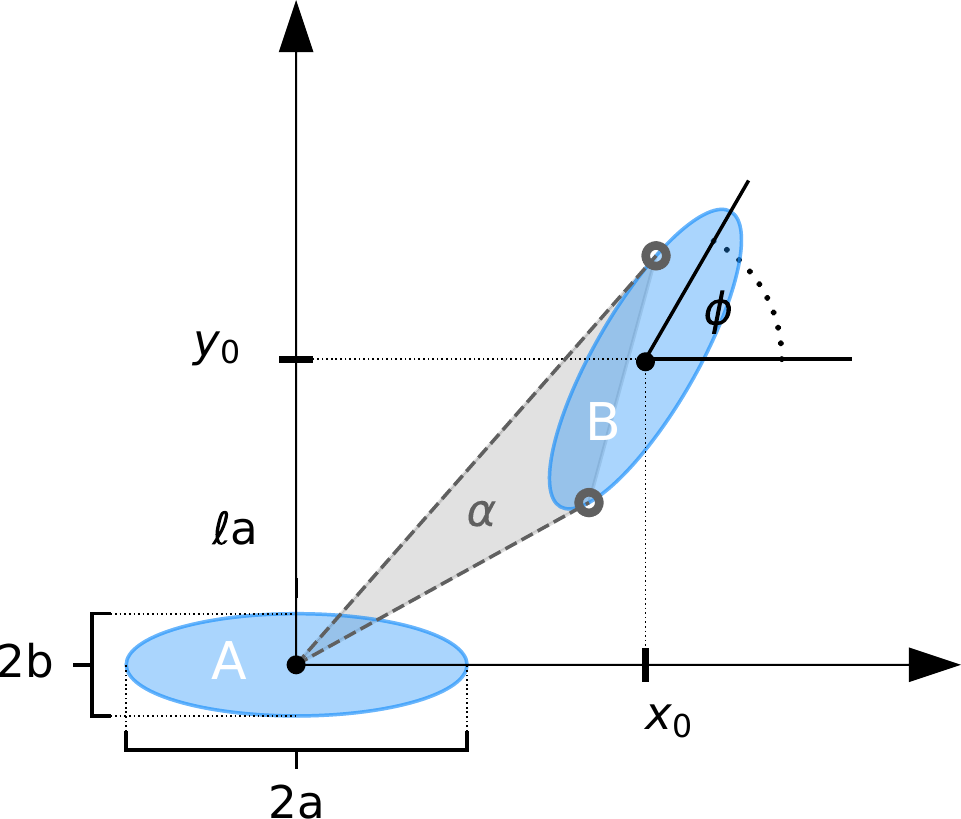}
    \caption{\textbf{Construction of visual field:} Illustration of the ellipses and variables used in the analytical calculation of the visual interactions. The visual angle, $\alpha$, of ellipse B in the visual field of ellipse A, is colored grey and given by the angle between the two tangent lines (dashed, grey) which intersect ellipse B in a single point (grey open circle) each.}
    \label{fig:sketch_ellipse}
\end{figure}

To construct the visual networks, we assume an ellipse at relative position $(x_0,y_0)=r(\cos\theta,\sin\theta)$, with major axis of length 1 rotated by $\phi$ from the x-axis and minor axis of length $w$ and calculate the angular area that it occupies in the visual field of an observer with 360$^\circ$ vision.
The observed ellipse is given by:
\begin{equation}
\begin{aligned}
 \left(\begin{array}{c}x\\y\end{array}\right)&=\left(\begin{array}{c} x_0+\frac{1}{2}\cos\psi\cos\phi+\frac{w}{2}\sin\psi\sin\phi\\
						     y_0+\frac{1}{2}\cos\psi\sin\phi-\frac{w}{2}\sin\psi\cos\phi\end{array}\right)
\label{eq:ellipse}	
\end{aligned}
\end{equation}
with
\begin{equation}
  0\leq \psi<2\pi~.
\end{equation}
We calculate the gradient of the observed ellipse as
\begin{equation}
\begin{aligned}
 \frac{dy}{dx}&=\frac{\sin\phi\sin\psi+w\cos\phi\cos\psi}{\cos\phi\sin\psi-w\sin\phi\cos\psi} ~.
 \end{aligned}
\end{equation}
A tangent line to the ellipse is given by
\begin{equation}
 y=\left.\frac{dy}{dx}\right|_s(x-x_s)+y_s
\end{equation}
where $(x_s,y_s)$ is the tangent point on the ellipse (see Fig. \ref{fig:sketch_ellipse}).
Since we place our observer at the origin, the tangents (dashed lines in Figure \ref{fig:sketch_ellipse}) we need in order to determine the visual field have to pass through this point and thus it needs to hold that
\begin{equation}
\begin{aligned}
0&=\left.-\frac{dy}{dx}\right|_sx_s+y_s\\
0&=-(dy/d\psi)|_sx_s+(dx/d\psi)|_sy_s~.
\end{aligned}
\end{equation}
Using \eqref{eq:ellipse} we can write
\begin{equation}
\begin{aligned}
       x_s&=r\cos\theta+\frac{\cos\phi\cos\psi_s+w\sin\phi\sin\psi_s}{2}\\
       \frac{dx}{d\psi}|_s&=\frac{-\cos\phi\sin\psi_s+w\sin\phi\cos\psi_s}{2}\\
       y_s&=r\sin\theta+\frac{\sin\phi\cos\psi_s-w\cos\phi\sin\psi_s}{2}\\
       \frac{dy}{d\psi}|_s&=-\frac{\sin\phi\sin\psi_s-w\cos\phi\cos\psi_s}{2}
\end{aligned}
\end{equation}
Solving for $\psi_s$ yields
\begin{equation}
\begin{aligned}
\psi_s^{\pm}=&\pm2\tan^{-1}\left(\frac{\gamma\mp r\sin(\theta - \phi)}{\beta}\right)\\
\beta=& w(2r\cos(\theta - \phi) - 1)\\
 \gamma=&\left(-w^2 + 2r^2\left((1 + w^2)+ (w^2 - 1)\cos(2(\theta-\phi))\right)\right)^{1/2}~.
\end{aligned}
\label{eq:psi_tps}
\end{equation}
Inserting \eqref{eq:psi_tps} into \eqref{eq:ellipse} returns the tangent points $(x_{x1}, y_{s1})$ and $(x_{x2}, y_{s2})$ whose polar angles, $\theta_{s1}$ and $\theta_{s1}$, determine the angular area of the ellipse, $\alpha=\text{min}(|\theta_{1s}-\theta_{2s}+n\pi|, n\in\mathbb{N})$ (here we use the fact that the angular area needs to be smaller than $\pi$ because the ellipses are not allowed to overlap).

Occlusions of individual $j$ in the visual field of $i$ by all other individuals $k\neq j$ are then determined by an algorithm using intersections of rays originating from the eye of the focal individual $i$ and going through the tangent points (as perceived by $i$) on $j$ with the outlines of ellipses $k$. To determine if $j$ or $k$ is visible to $i$ in the occluded area, intersection of ellipse outlines with the angle bisectors of the rays described above are also considered. 
\section{Equivalent visual and metric thresholds}
\begin{figure*}[htpb]
    \centering
    \includegraphics[width=.9\linewidth]{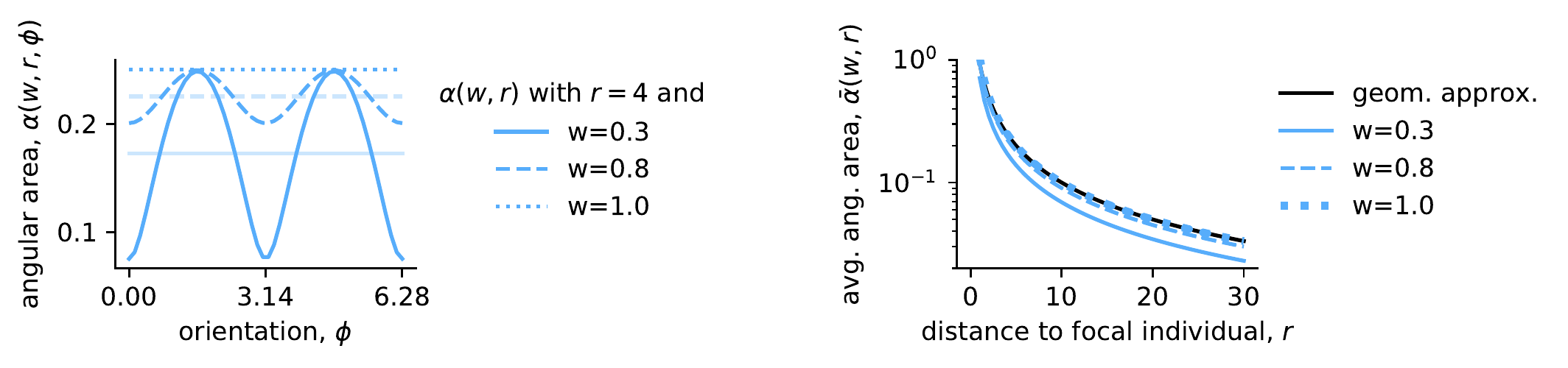}
    \caption{\textbf{Determining the equivalent visual threshold to a metric threshold:} \textbf{Left:} examples of angular areas, $\alpha(w,r,\phi)$ of an ellipse of length 1 and width $w=0.3$ (solid line), $w=0.8$ (dashed line) and $w=1.0$ (dotted line) at a distance of $r=4\,\text{BL}$ as a function of the relative orientation, $\phi$, of that ellipse with respect to its angular position in the visual field of the focal individual. The average angular area, $\bar{\alpha}(r,w)=\int_0^{2\pi}\alpha(r,w,\phi)d\phi$ is indicated by a horizontal line in the same line style. \textbf{Right:} The average angular area of an ellipse of random orientation and width $w$ as a function of distance, $r$, to the focal individual for the same example widths as on the left. A (commonly overestimating) approximation for the average angular area is given by $\bar{\alpha}(w,r)=2\arctan(\frac{1}{2r})$ (black line). The matching thresholds used in the main text for the comparison of the density dependence of the simple and complex contagion process on visual and metric networks were obtained from the numeric estimation of the average angular area for the specific value of $w$ depicted here by the blue lines.}
    \label{fig:L_eff}
\end{figure*}

For details on the construction of an equivalent visual and metric threshold refer to Figure \ref{fig:L_eff}.

\section{Infection probability}
\begin{figure*}[htpb]
    \centering
    \includegraphics[width=.6\linewidth]{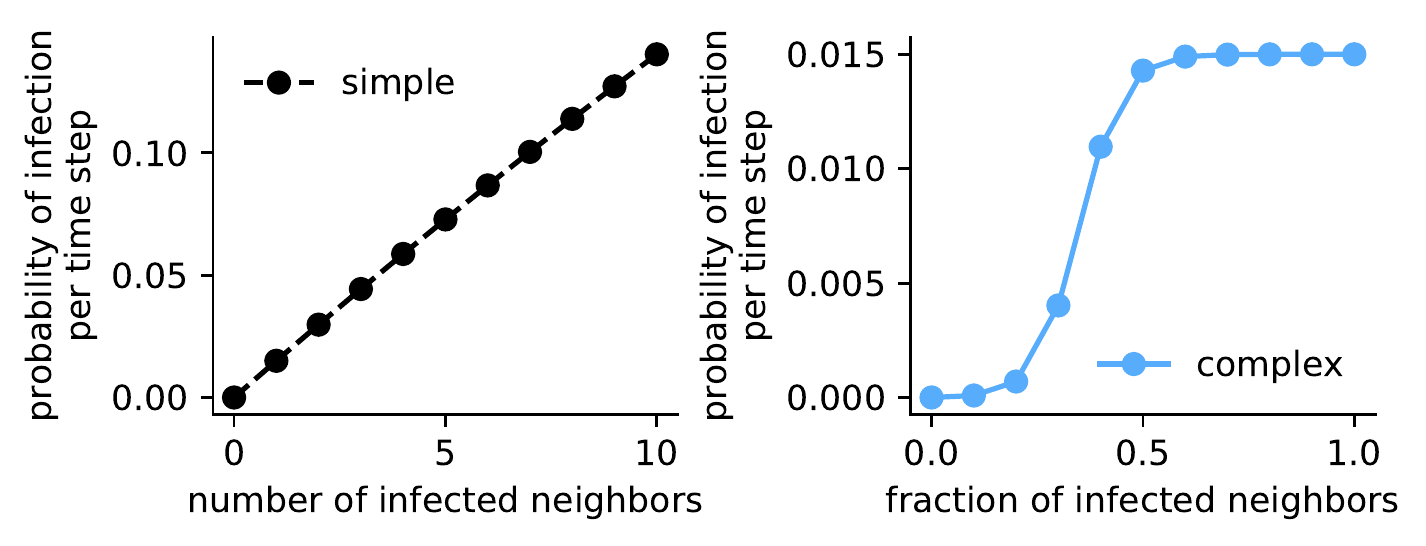}
    \caption{\textbf{Infection probability per time step of a single individual} (with degree 10) for the two contagion types. While for the simple contagion (left) each additional infected neighbor is adding similarly to the overall infection probability, for the fractional complex contagion (right), the influence of one additional infected neighbor is largest close to the transition at $r_0=0.35$. Parameters: $\Delta t=0.05, \beta=0.3, r_0=0.35, \sigma=10$.}
    \label{fig:infection_probability}
\end{figure*}

For an example illustrating the different infection probabilities for the two contagion types refer to Figure \ref{fig:infection_probability}.

\section{Link realization probability as function of link length}
To highlight the qualitative differences between the link length distributions of the different network types, which are not adequately captured by the average shown in the main text (Figure 2C), here we show histograms of the relative link lengths in a network for two densities and the same sets of thresholds. In grey we included the distribution of all possible links (i.e. the fully connected network).
To make the differences between networks more apparent, we also calculated the average fraction of realized links of a certain relative length (count of links of a certain length in a certain network type divided by all possible links of this length for this spatial configuration, i.e. the fully connected network).
 While in metric networks all links up to a (rescaled) threshold value are realized with a sharp cutoff at this value, topological networks include all links up to a certain length and then the distribution rapidly decays. In comparison to the other two types, the distribution of link lengths in visual networks has a slower decay towards long links. For all network types the shape of the distribution depends on the threshold value. Note that for low threshold values and low densities, the visual network has a non-vanishing probability for the realization of links of any length. 
\begin{figure*}[htpb]
    \centering
    \includegraphics[width=.9\linewidth]{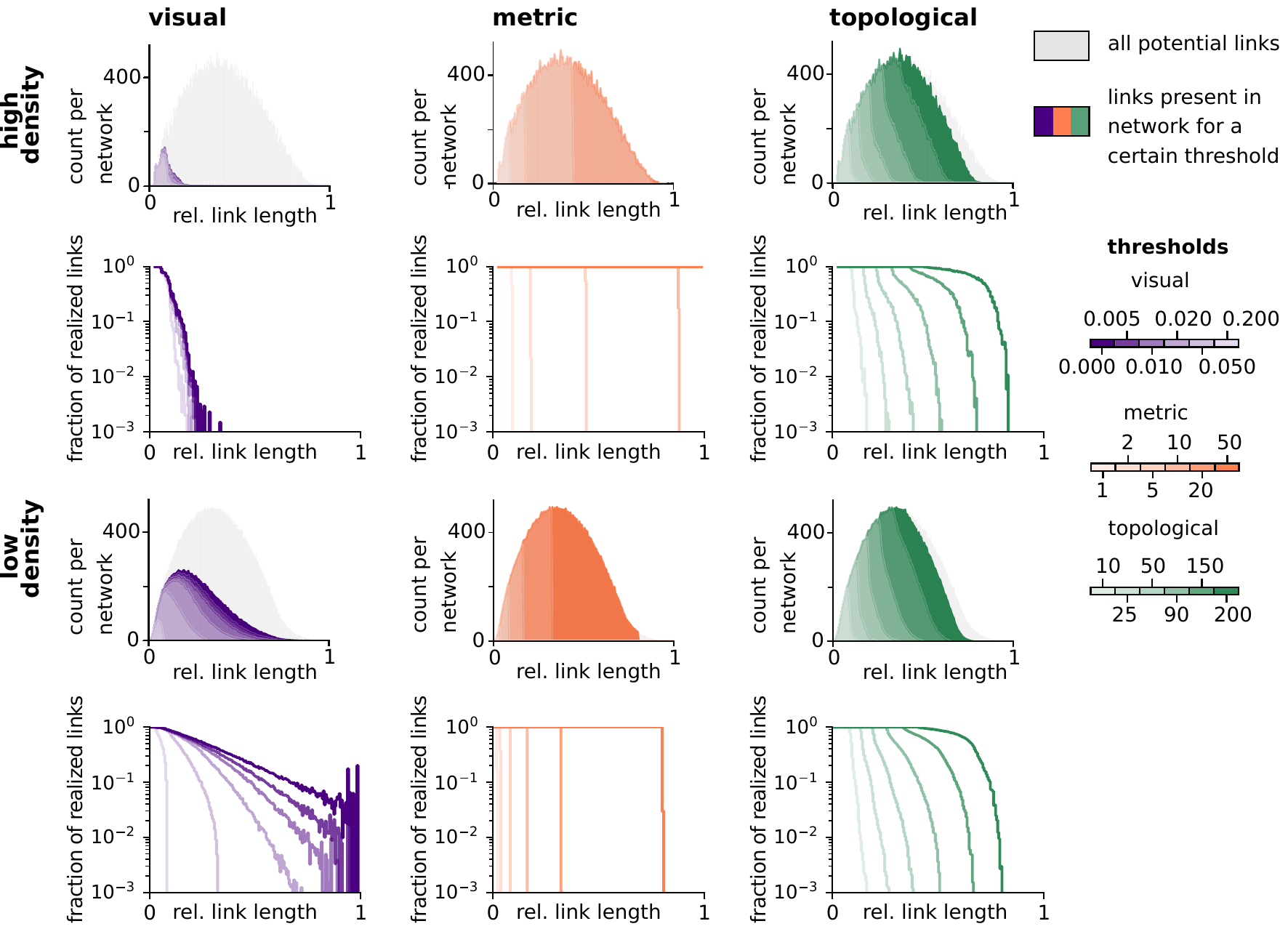}
    \caption{\textbf{The effect of density and network thresholds on the distribution of relative link lengths} for low (bottom two rows, $g=3.0$) and high (top two rows, $g=0.5$) density. The distributions of link length (colored according to threshold, legend as in (A) to (C)) shows a distinct shape for each network type. Of all the links in an all-to-all connected network (grey) only a certain part is present for the different network types depending on the threshold value. The fraction of realized links is the ratio of links of a certain length found in each network type and all possible links of this length in a fully connected network. 
    All subfigures were created using networks of N=225 and threshold values as in Figure 1 of the main text.}
    \label{fig:link_realization_prob}
\end{figure*}

\end{document}